\documentclass{ifacconf}

\pdfoutput=1
\usepackage[utf8]{inputenc}
\usepackage[pdftex]{graphicx}
\graphicspath{{./Figures}}
\usepackage{float}
\usepackage[font=footnotesize]{subcaption} 
\usepackage{caption}
\usepackage{amsmath, amssymb, amsfonts}
\usepackage{multirow}
\usepackage{natbib}

\usepackage{xcolor}

\makeatletter
\renewcommand\thesubsubsection{(\alph{subsubsection})}  
\makeatother

\newtheorem{remark}{Remark}

\usepackage[ruled, vlined, longend, linesnumbered, algo2e]{algorithm2e}
\SetKwInput{Input}{Input}

\newcommand{\R}{\mathbb{R}} 
\newcommand{\N}{\mathbb{N}} 
\newcommand{\Ni}[2]{\N_{#1}^{#2}} 
\newcommand{\cc}[1]{{\mathcal{#1}}} 
\newcommand{\set}[2]{\{#1 : #2\}}

\def\Sum#1#2{\sum\limits_{#1}^{#2}} 

\def\fracg#1#2{{\displaystyle{\frac{#1}{#2}}}} 
\def\vv#1{{ \rm \bf{#1}}} 

\def\tX{\vv{x}} 
\def\tU{\vv{u}} 
\def\nx{{n_x}} 
\def\nu{{n_u}} 
\def\nth{{n_\theta}} 
\def\traj{\cc{T}} 
\def\spaceT{\mathbb{T}} 
\def\cXo{\cc{X}_0} 
\def\spaceTo{\mathbb{T}(\cXo)} 

\def\prefFunc{\Pi} 
\newcommand{\surr}[1][\theta]{\pi_{#1}} 
\newcommand{\model}[1][\theta]{\sigma_{#1}} 
\newcommand{\probSurr}{P_{\surr[]}} 
\def\Pds{\cc{D}_{\prefFunc}} 
\def\np{n_p} 
\def\settime{{\kappa_{\varepsilon}}} 

\begin{document}
\begin{frontmatter}

\title{Active Learning MPC Objective Functions from Preferences}

\author{Hasna El Hasnaouy, Pablo Krupa, Mario Zanon, Alberto Bemporad}
\address{IMT School for Advanced Studies, Lucca, Italy}
\address{\small\{hasna.elhasnaouy, pablo.krupa, mario.zanon, alberto.bemporad\}@imtlucca.it}

\begin{abstract}
Designing the objective function in Model Predictive Control (MPC) is challenging when performance assessment criteria are available only from human judgment.
We adopt a preference-based learning (PbL) approach to learn the MPC objective function from preferences over trajectory pairs.
However, the real-world application of PbL is often restricted by the significant cost or limited availability of human preference queries.
To address this, Active Learning (AL) strategies seek to improve sampling efficiency, reducing the labeling effort required to obtain a well-performing classifier.
We present two AL strategies for learning the MPC objective function from human preferences over pairwise system trajectories: a pool-based strategy that selects trajectory pairs that are both uncertain under the current surrogate and diverse relative to previously labeled comparisons, and a query-synthesis strategy that incorporates new trajectories using the current surrogate-driven MPC.
Numerical results show that the proposed strategies yield closed-loop behaviors that align more with the expressed preference using fewer number of queries compared to a random sampling approach.
\end{abstract}


\end{frontmatter}
\section{Introduction}

Model Predictive Control (MPC) is an advanced control strategy where the  control input is computed by solving an optimization problem at each time step~\citep{Rawlings_MPC_2017}. 
A crucial part of MPC design is the choice of the objective function, as it determines the closed-loop behavior of the system. 
However, choosing this function is challenging, as there is often no immediate relation between its parameters and the resulting closed-loop behavior under the receding horizon mechanism.
Therefore, the calibration of MPC parameters is a demanding process that may require prior experience or repeated fine-tuning to obtain the desired closed-loop behavior~\citep{Garriga2010ModelPC}.
Several MPC tuning approaches have been proposed in the literature, including the use of global optimization techniques~\citep{Brochu2010, Forgione2020} or reinforcement learning~\citep{Gros2020}. 
However, these methods rely on the availability of an explicit performance metric that quantifies the desired closed-loop behavior. 

When explicit performance metrics are unavailable or difficult to formalize, preference-based learning (PbL) provides an alternative relying instead on qualitative human judgments to guide the learning process~\citep{Wirth_JMLR_2017}. 
This paradigm has been successfully applied in various domains, including deep reinforcement learning from pairwise preferences \citep{christiano2017deep}, training large language models~\citep{Ziegler2019FineTuningLM, Stiennon2020} and controller tuning~\citep{Coutinho2024, Zhu_ECC_2021, Zhu_TCST_2022}.
In our recent work~\citep{Krupa_CDC_25}, which we extend in this paper, we proposed a PbL approach for learning the MPC objective function  from human preferences expressed over pairs of closed-loop trajectories.
By formulating the problem as a binary classification task solved by logistic regression, we showed that a surrogate model can be trained to capture the underlying preference function, and this model can then be directly used as the MPC objective function to achieve a closed-loop behavior that aligns with the human preference.

Although PbL enables us to learn from qualitative assessments, its practical application is often restricted by the significant cost of acquiring human preferences. 
In our setting, each human query requires comparing two system trajectories and providing a label indicating which one is preferred, a process that is time-consuming.
To address the high cost of labeling data, Active Learning (AL) strategies were introduced to increase the learning rate by strategically selecting the most informative samples for labeling, rather than random sampling~\citep{cohn1994,prince2004}.
Several approaches for AL have been proposed, including pool-based sampling, which selects instances from a static unlabeled dataset, stream-based sampling, which selects and then labels instances from a stream, and query synthesis, which actively generates new instances for labeling~\citep{settles2009active, Bemporad_IS_2023}.
These approaches have been applied across diverse domains, including natural language processing~\citep{schumann2019active}, image classification~\citep{gal2017}, and control applications~\citep{schillinger2017safe,saviolo2023active}.
For logistic regression, both experimental design methods and uncertainty-based heuristics have been explored~\citep{schein2007active}.

This paper introduces two active learning strategies to enhance the learning rate of the preference-based MPC tuning presented in~\cite{Krupa_CDC_25}.
The first is a pool-based approach that selects trajectory pairs by combining uncertainty-based and diversity-aware criteria, where we include an intra-pair diversity term specific to the pairwise-preference setting. The second is a query-synthesis strategy that generates new trajectory pairs during training by exploring diverse initial conditions and comparing trajectories from random and learned MPC controllers. Both strategies aim to achieve a closed-loop performance that aligns with the human preference with fewer queries than random sampling.

The rest of this paper is organized as follows.
Section~\ref{sec:problem} reviews the preference-based learning framework for MPC tuning from pairwise trajectory comparisons presented in~\cite{Krupa_CDC_25}.
Section~\ref{sec:ALstrategies} presents the proposed pool-based and query-synthesis AL strategies.
Section~\ref{sec:results} provides numerical results on a system of oscillating masses, comparing the proposed AL strategies against random sampling.
Finally, Section~\ref{sec:conclusions} concludes the paper.

\section{Learning the MPC objective function from pairwise preferences} \label{sec:problem}

We consider a discrete-time dynamical system of the form:
\begin{equation} \label{eq:sys}
    x(t+1) = f(x(t), u(t)), \\
\end{equation}
where $x(t) \in \R^\nx$ denotes the system state and $u(t) \in \R^\nu$ the control input at time step $t$.  
The control objective is to regulate the system to the origin, which is assumed to be an equilibrium point, i.e., $f(0,0)=0$, while satisfying system constraints $x(t) \in \cc{X}$, $u(t) \in \cc{U}$, $\forall t$, where $\cc{X} \subseteq \R^\nx$ and $\cc{U} \subseteq \R^\nu$ are closed sets containing the origin in their interiors.
We want to satisfy this control objective for any initial state $x(0) \in \cXo$, where the set of initial states $\cXo \subseteq \cc{X}$ is compact.

For a given horizon $N > 0$, let $\spaceT$ denote the space of trajectories $\tX \doteq \{x_i\}_{i=0}^{N}$ and $\tU \doteq \{u_i\}_{i=0}^{N-1}$ that satisfy the system dynamics and constraints, i.e.,
\begin{equation*}
    \spaceT = \set{(\tX, \tU)}{ x_i \in \cc{X}, u_i \in \cc{U}, \; x_{i+1} = f(x_i, u_i)}.
\end{equation*}
We denote by $\spaceT(\hat{x}) \doteq \{(x,u)\in \spaceT \mid x_0 = \hat{x}\}$ the subset of trajectories starting from a given initial condition $\hat{x} \in \R^\nx$, and by $\spaceTo$ the subset of $\spaceT$ with initial conditions in $\cXo$.

We satisfy the control objective using an MPC controller whose optimization problem is given by
\begin{subequations} \label{eq:prefMPC}
\begin{align}
    \min\limits_{\traj \doteq (\tX, \tU)} \;& \model(\traj) \label{eq:icnnMPC:cost} \\
    {\rm s.t.} \quad & x_0 = x(t), \label{eq:prefMPC:init} \\
    & x_{i+1} = f(x_i, u_i), \quad i \in \Ni{0}{N-1}, \label{eq:prefMPC:dynamics} \\
    & x_i \in \cc{X}, \quad i \in \Ni{1}{N-1}, \label{eq:prefMPC:constr:x} \\
    & u_i \in \cc{U}, \quad i \in \Ni{0}{N-1}, \label{eq:prefMPC:constr:u}
\end{align}
\end{subequations}
for some continuous objective function $\model \colon \spaceT \to \R$ parameterized by $\theta \in \R^\nth$.
The classical MPC objective function takes the form
\begin{equation} \label{eq:quad_cost}
    \model = \| x_N \|_P^2 + \sum_{i = 0}^{N-1} \| x_i \|_Q^2 + \| u_i \|_R^2,
\end{equation}
for some choice of positive definite matrices $Q, P \in \R^{\nx \times \nx}$ and $R \in \R^{\nu \times \nu}$.
In~\eqref{eq:quad_cost}, we have $\theta = (Q, R, P)$.
Let $\vv{x}^*(t)$ and $\vv{u}^*(t)$ be the optimal solution of~\eqref{eq:prefMPC} for a given $x(t) \in \cc{X}$ (assuming that~\eqref{eq:prefMPC} is feasible for this $x(t)$).
The control law of~\eqref{eq:prefMPC} is $u(t) = u_0^*(t)$, where $u_0^*(t)$ is the first control input in the sequence $\vv{u}^*(t)$.
Conditions under which the MPC law is stabilizing and provides constraint satisfaction are well known in the control literature, see, e.g., \cite{Rawlings_MPC_2017}.\footnote{Stabilizing ingredients typically include a suitable terminal set $\cc{X}_f \subseteq \cc{X}$ and terminal constraint $x_N \in \cc{X}_f$, though this can also be achieved without terminal constraints under certain conditions \citep{grne2013nonlinear}. We don't include a terminal constraint in~\eqref{eq:prefMPC} for simplicity of exposure, as our focus is on learning $\model$.}

The challenge considered in this paper is to learn the parameters $\theta$ of the objective function $\model$ such that the transient behavior of the closed-loop system~\eqref{eq:sys}-\eqref{eq:prefMPC} matches a human judgment of ``desirable'' behavior, or performance.
However, this performance criterion is not available in explicit form.
Instead, it is only revealed indirectly by querying the human over their preference between a pair of trajectories.
Human preferences are captured by a binary preference function $\prefFunc \colon \spaceT \times \spaceT \to \{0,1\}$, which compares two candidate trajectories $\traj,\cc{S} \in \spaceT$:
\begin{equation} \label{eq:prefFunc}
\prefFunc(\traj, \cc{S}) =
\begin{cases}
    1, \; \text{if $\traj$ is judged preferable to $\cc{S}$}, \\
    0, \; \text{otherwise}.
\end{cases}
\end{equation}

Let $\Pds$ be an available dataset of $\np$ trajectory pairs along with their associated preference labels, i.e.,
\begin{equation} \label{eq:dataset}
    \Pds = \{(\traj_\ell, \cc{S}_\ell, p_\ell)\}_{\ell=1}^{\np},  
\end{equation}
where $p_\ell \doteq \prefFunc(\traj_\ell, \cc{S}_\ell)$ and each $\traj_\ell, \cc{S}_\ell \in \spaceT(x_\ell)$ for some initial state $x_\ell \in \cXo$.
In~\cite{Krupa_CDC_25}, the authors propose the use of a surrogate model $\surr \colon \spaceT \times \spaceT \to \{0,1\}$, whose parameters $\theta$ are trained by solving a binary classification problem maximizing the fit between $\surr$ and $\prefFunc$ for the trajectory pairs and preferences in $\Pds$.
In particular, the surrogate model $\surr$ is taken as 
\begin{equation} \label{eq:surrogate}
\surr(\traj, \cc{S}) =
\begin{cases}
    1 \; \text{if $\model(\traj) \leq \model(\cc{S})$}, \\
    0 \; \text{if $\model(\traj) > \model(\cc{S})$}.
\end{cases}
\end{equation}
That is, the surrogate model predicts the preferred trajectory as the one providing the smaller value of $\model$.
As discussed in~\cite{Krupa_CDC_25}, the choice of~\eqref{eq:surrogate} is taken so that the use of $\model$ as the objective function in~\eqref{eq:prefMPC} provides a closed-loop behavior that aligns with the human preference, assuming that a good fit of the underlying preference function $\prefFunc$ has been achieved.

As typically done in binary classification, the output of the surrogate model~\eqref{eq:surrogate} is modeled as a probability using the sigmoid function
\begin{equation} \label{eq:prob}
    \probSurr (\traj, \cc{S}) = \fracg{1}{1 + \exp(\model(\traj) - \model(\cc{S}))},
\end{equation}
providing $\probSurr (\traj, \cc{S}) > 0.5$ if $\model(\traj) < \model(\cc{S})$, and vice versa.
Note that $\probSurr \simeq 1$ if $\model(\traj) \ll \model(\cc{S})$, $\probSurr \simeq 0$ if $\model(\traj) \gg \model(\cc{S})$, and $\probSurr \simeq 0.5$ if $\model(\traj) \simeq \model(\cc{S})$; this last case reflecting a high uncertainty between trajectories providing similar values of $\model$.

Using the probability model~\eqref{eq:prob}, the learning problem for a given dataset $\Pds$ of size $\np$ is given by
\begin{equation} \label{eq:learn:prob}
    \min\limits_\theta r(\theta) + \frac{1}{\np} \Sum{\ell=1}{\np} \cc{L}(p_\ell, \probSurr(\traj_\ell, \cc{S}_\ell)),
\end{equation}
where $\cc{L}\colon \R \times \R \to \R$ is the cross-entropy loss
\begin{equation*} \label{eq:learn:func:cross}
        \cc{L}(p, \hat{p}) = -p \log(\hat{p}) - (1 - p) \log(1 - \hat{p}),
\end{equation*}
which penalizes mismatches between a predicted value $\hat{p} \in [0, 1]$ and the observed label $p \in \{0, 1\}$, ensuring that the surrogate aligns as closely as possible with the preferences collected in $\Pds$.
Function $r \colon \R^\nth \to \R$ is a regularization term, typically an $\ell_1$ or $\ell_2$ norm, e.g., $r(\theta) = \rho \| \theta \|_2^2$, for some $\rho > 0$.
Problem~\eqref{eq:learn:prob} can be solved using standard machine learning tools, such as Adam~\citep{kingma_adam_2014} or L-BFGS-B~\citep{byrd_LBFGS_JSC_1995, bemporad2025bfgs}.

As shown in~\cite{Krupa_CDC_25}, using the function $\model$ obtained from solving~\eqref{eq:learn:prob} as the objective function of the MPC~\eqref{eq:prefMPC} can provide a closed-loop behavior that aligns with the collected human preference.
However, obtaining this result may require a large dataset $\Pds$, which in turn requires performing a large amount of human queries.\footnote{We note that in~\cite{Krupa_CDC_25}, learning models $\model$ that outperformed the controllers used to generate the trajectories in $\Pds$ required hundreds of queries.}
This can be time consuming and/or expensive.
In the following section we present two AL strategies whose objectives are to reduce the number of queries required to obtain a well-performing MPC controller, as measured by its alignment with the collected preferences.

\begin{remark}
Note that the dataset~\eqref{eq:dataset} only pairs two trajectories $(\traj, \cc{S})$ if both share the same initial condition $x_0 \in \cXo$, i.e., if $\traj, \cc{S} \in \spaceT(x_0)$.
The reason is that comparisons between trajectories are only meaningful if they start from the same initial condition.
To illustrate this, consider a trajectory $\hat{\traj} \in \spaceT(\hat{x})$, with $\hat{x} \in \cXo$, $\hat{x} \neq 0$.
Suppose the human consistently prefers $\hat{\traj}$ over all other trajectories starting from $\hat{x}$, i.e., $\prefFunc(\hat{\traj}, \traj) = 1$ $\forall \traj \in \spaceT(\hat{x})$, $\traj \neq \hat{\traj}$. 
Now, if we compare $\hat{\traj}$ with the trivial trajectory $\traj_0$, defined by $x_i = 0$ for all $i \in \Ni{0}{N}$ and $u_i = 0$ for all $i \in \Ni{0}{N-1}$, the human would naturally prefer $\traj_0$, as it perfectly achieves the control objective of reaching the origin. 
However, this comparison does not reveal any useful information about how \emph{good} $\hat{\traj}$ is relative to other trajectories $\traj \in \spaceT(\hat{x})$.  
\end{remark}

{
\setlength{\algomargin}{1.0em}
\begin{algorithm2e}[t]
    \DontPrintSemicolon
    \caption{General AL loop} 
    \label{alg:ALloop}
    \Input{Initial labeled dataset $\Pds^0$}
    Learn initial $\theta^0$ by solving~\eqref{eq:learn:prob} using $\Pds^0$.\;
    \For{$k = 1, 2, \dots$, M}{
        Obtain new trajectory pairs $\{(\traj_i, \cc{S}_i)\}_{i = 1}^{n_k}$\;
        Query the human for $p_i = \prefFunc(\traj_i, \cc{S}_i)$, $i \in \Ni{1}{n_k}$\;
        $\Pds^k = \Pds^{k-1} \cup \{(\traj_i, \cc{S}_i, p_i)\}$\;
        Learn $\theta^k$ by solving~\eqref{eq:learn:prob} using $\Pds^k$.
    }
\end{algorithm2e}
}

{
\setlength{\algomargin}{1.0em}
\begin{algorithm2e}[t]
      \DontPrintSemicolon
      \caption{Pool-Based AL strategy}
      \label{alg:poolAL}
      \Input{$\Pds^0$, unlabeled pool $\cc{Q}$, batch size $n_k > 0$}
      Learn initial $\theta^0$ by solving~\eqref{eq:learn:prob} using $\Pds^0$\;
      \For{$k = 1, 2, \dots, M$}{
          $a(\mathcal{P}) = U(\cc{P}) D^k(\cc{P})$ for all $\cc{P} \in \cc{Q}$\;
          Select $\{(\traj_i, \cc{S}_i)\}_{i=1}^{n_k}$ as top $n_k$ pairs by $a(\cc{P})$\;
          Query the human for $p_i = \prefFunc(\traj_i, \cc{S}_i)$, $i \in \Ni{1}{n_k}$\;
          $\Pds^k = \Pds^{k-1} \cup \{(\traj_i, \cc{S}_i, p_i)\}_{i=1}^{n_k}$\;
          $\cc{Q} = \cc{Q} \setminus \{(\traj_i, \cc{S}_i)\}_{i=1}^{n_k}$\;
          Learn $\theta^k$ by solving~\eqref{eq:learn:prob} using $\Pds^k$\;
      }
  \end{algorithm2e}
}

\section{Active learning strategies}~\label{sec:ALstrategies}

\vspace*{-2em}
The goal of AL is to minimize the number of samples required to learn a good quality model.
Starting with an initial dataset~\eqref{eq:dataset}, which we label as $\Pds^0$, at each iteration $k$ of the AL loop, we acquire $n_k$ new trajectory pairs, query the human for their preference between them, add them to the labeled dataset, and finally retrain the surrogate model using the updated dataset $\Pds^k$.
This process is summarized in Algorithm~\ref{alg:ALloop}.
The AL loop is performed until a stopping criteria is satisfied.
In Algorithm~\ref{alg:ALloop} we consider performing a fixed number of iterations $M > 0$.
However, other stopping criteria are possible~\citep[\S 3.4]{Ren_deelAL_survey_21}.
In the following subsections, we present two AL strategies for the preference-based learning of MPC objective function presented in Section~\ref{sec:problem}.

\subsection{Pool-based Active Learning} \label{sec:poolAL}

We present a pool-based AL strategy where we assume that we have a pool of unlabeled trajectory pairs and we need to decide which one to present to the human.
Unlike random sampling, which treats all unlabeled pairs equally, this method prioritizes queries based on their expected contribution to learning, as measured by an acquisition function, where pairs with the highest acquisition values are considered the most informative to query.

Given an initial labeled dataset $\Pds^0$ and an unlabeled pool $\cc{Q} = \{ \cc{P}_i\}_{i = 1}^{n_q}$ of candidate trajectory pairs $ \cc{P}_i = (\traj_i, \cc{S}_i)$, the algorithm iteratively selects batches of pairs to query, as presented in Algorithm~\ref{alg:poolAL}. At each iteration $k$ of the AL strategy, $n_k > 0$ new pairs from $\cc{Q}$ are added to $\Pds^k$ by considering two metrics: uncertainty of the surrogate model and trajectory diversity.

For a given trajectory pair $\cc{P} = (\traj_1,\traj_2)$ with $\traj_1,\traj_2 \in \spaceT$, uncertainty is measured as
\begin{equation} \label{eq:uncertainty}
    U(\cc{P}) \doteq 0.5 - \left| \probSurr(\traj_1,\traj_2) - 0.5 \right|.
\end{equation}
This score is highest when the probability $\probSurr(\traj_1,\traj_2)$ is close to $0.5$, see~\eqref{eq:prob}, i.e., when the surrogate model is uncertain about which trajectory is preferred, whereas values close to $0$ correspond to low uncertainty.

The common approach for measuring diversity in the AL literature is to maximize (in some sense) the distance between the existing labeled samples and the new query-point.
In our setting, we take the distance metric between two trajectory pairs $\cc{P}_\traj = (\traj_1,\traj_2)$ and $\cc{P}_\cc{S} = (\cc{S}_1,\cc{S}_2)$ as
\begin{equation*}\label{eq:distance_pair}
d_p(\mathcal{P}_\traj,\mathcal{P}_\mathcal{S})
{\doteq} \min\{ d(\traj_1,\mathcal{S}_1) + d(\traj_2,\mathcal{S}_2),
         d(\traj_1,\mathcal{S}_2) + d(\traj_2,\mathcal{S}_1) \}
\end{equation*}
where $d\colon \spaceT \times \spaceT \to \R$ is a distance between trajectories, such as the Euclidean distance
\begin{equation} \label{eq:euclidean_traj_dist}
    d(\traj_1,\traj_2) \doteq \sum_{i=0}^{N} \|x_i^1-x_i^2\|_2^2 + \sum_{i=0}^{N-1} \|u_i^1-u_i^2\|_2^2.
\end{equation}
Distance $d_p(\mathcal{P}_\traj,\mathcal{P}_\mathcal{S})$ is $0$ if the trajectories in $\mathcal{P}_\traj$ are the same as the ones in $\mathcal{P}_\mathcal{S}$, and increases with the distance between trajectories.
Using $d_p$, we define the diversity of a candidate pair $\cc{P}$ with respect to $\Pds^k$ as its minimum distance to any labeled pair in $\Pds^k$:
\begin{equation} \label{eq:diversity}
    D^k_p(\cc{P}) = \min_{\cc{P}_\traj \in \Pds^{k}} d_p(\cc{P}_\traj, \cc{P}).
\end{equation}
We find that only using $D^k_p$ generally provides poor results, as it only measures diversity between the trajectories in the candidate pair $\cc{P} = (\traj_1,\traj_2)$ with respect to the trajectories from the labeled pairs in $\Pds^k$, but does not take into account the intra-pair diversity between trajectories $\traj_1$ and $\traj_2$.
Note that, as discussed in Section~\ref{sec:problem}, a trajectory pair $\cc{P} = (\traj_1,\traj_2)$ where $d(\traj_1,\traj_2) \simeq 0$ will always have a high uncertainty, as $\model(\traj_1) \simeq \model(\traj_2)$; see~\eqref{eq:prob}.
Therefore, we propose to also account for diversity between the trajectories of a candidate $\cc{P}$ when measuring diversity, leading to the acquisition function
\begin{equation} \label{eq:acquisition}
    a(\cc{P}) = U(\cc{P}) D^k(\cc{P}),
\end{equation}
where $D^k(\cc{P})$ combines both $D^k_p(\cc{P})$ and $d(\cc{P}) \doteq d(\traj_1,\traj_2)$.
$D^k(\cc{P}) = D^k_p(\cc{P}) + d(\cc{P})$ yields the most consistent
performance, although other alternatives also work well.
We find that $D^k(\cc{P}) = D^k_p(\cc{P}) + d(\cc{P})$ provides the most consistent results, although other alternatives are possible, cf. Fig.~2.

The pool-based strategy selects the $n_k$ pairs with the highest acquisition scores~\eqref{eq:acquisition} from $\mathcal{Q}$.
These selected pairs are queried by the preference function $\prefFunc$, added to the labeled dataset $\Pds^k$, and removed from $\mathcal{Q}$.
The surrogate model is then retrained by solving~\eqref{eq:learn:prob} using the new $\Pds^k$.

\subsection{Query-synthesis active learning} \label{sec:queryAL}

{
\setlength{\algomargin}{1.0em}
\begin{algorithm2e}[t]
    \DontPrintSemicolon
    \caption{Query-synthesis AL strategy} 
    \label{alg:queryAL}
    \Input{Initial $\Pds^0$ and pool of controllers $\cc{C}^0$.}
    Learn initial $\theta^0$ by solving~\eqref{eq:learn:prob} using $\Pds^0$.\;
    \For{$k = 1, 2, \dots$}{
        Select new initial state $x_0^k \in \cXo$, e.g., using~\eqref{eq:greedy_dist}.\;
        Pick rand. $\kappa_i \in \cc{C}^{k-1}$. Gen. $\traj_i \in \spaceT(x^k_0)$ using $\kappa_i$.\;
        Gen. $\traj_j\in \spaceT(x^k_0)$ from MPC~\eqref{eq:prefMPC} with $\theta^{k-1}$.\;
        $\Pds^k = \Pds^{k-1} \cup \{(\traj_i, \traj_j, \prefFunc(\traj_i, \traj_j))\}$\;
        Learn $\theta^k$ by solving~\eqref{eq:learn:prob} using $\Pds^k$.\;
        $\cc{C}^k = \cc{C}^{k-1} \cup \{\text{Control law of MPC~\eqref{eq:prefMPC} with $\theta^k$}\}$
    }
\end{algorithm2e}
}

We present an AL strategy that is based on synthesizing new trajectory pairs online and adding them to $\Pds^k$.
We assume that the initial trajectories in $\Pds^0$ are generated using an initial pool of $n_\kappa$ controllers $\cc{C}^0 = \{\kappa_i\}_{i=1}^{n_\kappa}$, where each $\kappa_i \colon \R^\nx \to \R^\nu$ is a state-feedback control law.

The query-synthesis strategy is summarized in Algorithm~\ref{alg:queryAL}.
At each iteration $k$ of the Al loop, we first select a new initial state $x^k_0 \in \cXo$.
The selection of $x^k_0$ should be based on exploration from existing samples.
Let $\bar{\cc{X}}_0^k = \{x^k_{0,\ell}\}_{\ell=1}^{\np}$ be the collection of all the initial states of the trajectories in the dataset $\Pds^k$.
Then, $x^k_0$ can be taken, for instance, by using a greedy method~\citep{Yu_ICDM_2010}%
\begin{subequations} \label{eq:greedy_dist}
\begin{align}
    x_0^k &= \arg\max_{x \in \cXo} d_x(x), \\
    d_x(x) &= \min_{\ell = 1, \ldots, \np} \| x - x^{k-1}_{0,\ell} \|_2^2,
\end{align}
\end{subequations}
where $x_{0, \ell}^{k-1} \in \bar{\cc{X}}_0^{k-1}$.
The query-synthesis strategy then generates two closed-loop trajectories starting from $x_0^k$.
One trajectory is generated using the MPC~\eqref{eq:prefMPC} with $\theta^{k-1}$, and the other is generated using a random choice of the controllers in $\cc{C}^{k-1}$.
Both trajectories are added to $\Pds^k$ along with the query of the preference between them, and then used to retrain the surrogate model.
Finally, we augment $\cc{C}^k$ by adding the new learned MPC controller.

The idea of this AL strategy is to synthesize new trajectories using the current $\model$ and compare them with alternative controllers, including MPC controllers leaned in previous iterations of the AL loop.
As the size of the dataset increases, the learned MPC is expected to become better at providing a closed-loop behavior that aligns with the human preference.
If this is the case, then the preference between the newly added pair of trajectories should be the one generated by the latest value of $\theta^k$.
We note that this AL strategy lacks an exploration term, and is therefore prone to getting stuck in some solution, although we find it often outperforms the controllers in $\cc{C}^0$, see Section~\ref{sec:results}.

\section{Numerical results}\label{sec:results}

We illustrate the proposed framework on the system of oscillating masses used in~\citep[\S III]{Krupa_CDC_25}, which consists of three objects connected by springs. 
The state vector is $x = (p_1, p_2, p_3, v_1, v_2, v_3)$, where $p_i$ and $v_i$ denote the position and velocity of each object, respectively. 
The input vector is $u = (F_1, F_2)$, representing the forces applied to the two outermost objects, while the output is taken as $y = (p_1, p_2, p_3)$. 
The dynamics are given by a linear state-space model
\begin{subequations} \label{eq:lin_sys}
\begin{align}
    x(t+1) &= A x(t) + B u(t), \label{eq:lin_sys:x} \\
    y(t) &= C x(t), \label{eq:lin_sys:y}
\end{align}    
\end{subequations}
with the control objective of steering the system to $(x, u) = (0, 0)$ while satisfying the input constraints $|u(t)| \leq 2$.

We consider the human preference criteria used in~\cite[\S III.B]{Krupa_CDC_25}, which selects the trajectory with the smallest output settling time, measured as\footnote{We take $\settime(\traj) = N$ if $\traj$ does not converge within $N$ sample times.}
\begin{equation*}
    \settime(\traj) = \min\limits_{i \geq 0} k \colon i \in \N, \| y_i \| \leq \varepsilon, \| y_\ell \| \not> \varepsilon\; \forall \ell > i,
\end{equation*}
where $y_i = C x_i$ and we take $\varepsilon = 0.05$.
The preference function~\eqref{eq:prefFunc} is given by
\begin{subequations} \label{eq:results:complex:prefFunc}
\begin{equation} 
\prefFunc_\settime(\traj, \cc{S}) =
\begin{cases}
    1\; \text{if}\; \settime(\traj) < \settime(\cc{S}), \\
    0\; \text{if}\; \settime(\traj) > \settime(\cc{S}), \\
    \prefFunc_u(\traj, \cc{S}) \; \text{otherwise},
\end{cases}
\end{equation}
where, denoting $\max \| \vv{u}(\traj) \|_\infty$ as the maximum control action in $\traj$ (as measured by the infinity norm):
\begin{equation}
\prefFunc_u(\traj, \cc{S}) {=}
\begin{cases}
    1\; \text{if}\; \max \|\vv{u}(\traj)\|_\infty {\leq} \max \|\vv{u}(\cc{S})\|_\infty \\
    0\; \text{otherwise}.
\end{cases}
\end{equation}
\end{subequations}
That is, if $\traj$ and $\cc{S}$ have the same settling time, the one with the smallest maximum control action is preferred.

To solve problem~\eqref{eq:learn:prob}, we first run $1000$ iterations of Adam (taking $\beta_1 = 0.9$, $\beta_2 = 0.999$ and the learning rate as $0.001$) and then run the \texttt{jaxopt} L-BFGS-B solver~\citep{jaxopt_implicit_diff} starting from the solution provided by Adam.
We take $r(\theta) = 10^{-4} \| \theta \|_2^2$.
We evaluate the performance of an MPC~\eqref{eq:prefMPC} by using a fixed set of random initial states $\hat{\vv{x}} = \{\hat{x}_i\}_{i=1}^{10}$, with each $\hat{x}_i \in \cXo$.
Performance is evaluated by performing a closed-loop simulation starting from each $\hat{x}_i$ and measuring the values of $\settime(\hat{\traj}_i)$ and $\max \| \vv{u}(\hat{\traj}_i) \|_\infty$ for the resulting closed-loop trajectory $\hat{\traj}_i$.
At each AL iteration, problem~\eqref{eq:learn:prob} is solved twice, once starting from a random initialization of $\theta$, and once starting from $\theta^{k-1}$.
We keep the $\theta$ providing the smallest maximum value of $\settime$ in the performance evaluations using $\hat{\vv{x}}$.
In a practical setting, this would require querying the human for their preference over the two MPC controllers obtained from solving~\eqref{eq:learn:prob} at each iteration of the AL loop.

We take $\cXo = \{x \colon | p_i | \leq 0.2, | v_i | \leq 0.05, i \in \Ni{1}{3}\}$ and generate the initial pool of controllers $\cc{C}^0$ by taking $10$ MPC controllers~\eqref{eq:prefMPC} with $N = 30$ and a quadratic objective function~\eqref{eq:quad_cost} with random diagonal matrices $Q$, $R$ and $P$.
We generate the initial dataset $\Pds^0$ by taking $20$ states sampled from $\cXo$ using a uniform distribution for each element.
For each initial state, we select a random choice of two distinct controllers from $\cc{C}^0$ to generate two closed-loop trajectories, which are added to $\Pds^0$ along with the results of the preference query.
For the pool-based AL strategy, we take $n_k = 1$ and generate an unlabeled pool $\cc{Q}$ of $280$ trajectory pairs using the same $\cc{C}^0$ and method taken for $\Pds^0$.

\begin{figure*}[t]
    \centering
    \begin{minipage}{0.49\textwidth}
        \includegraphics[width=\linewidth]{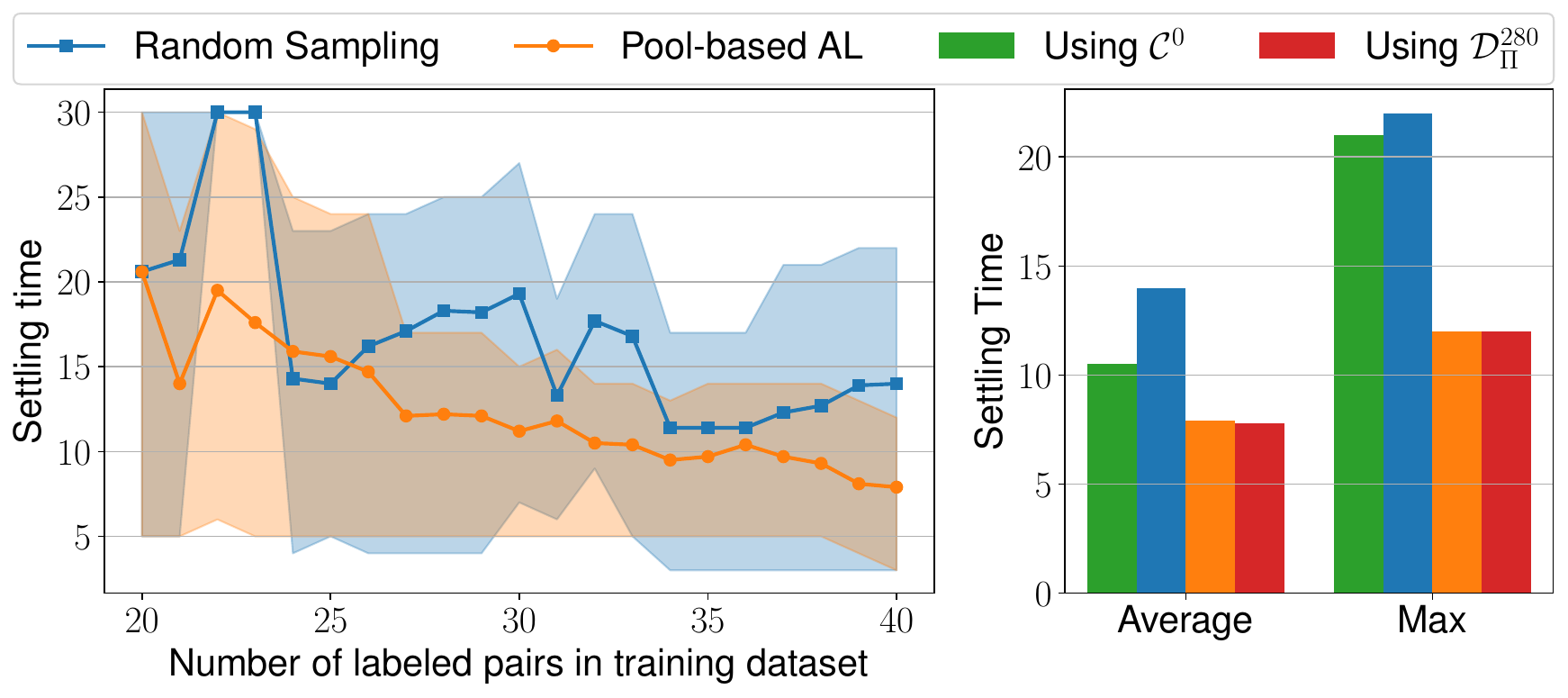}
    \end{minipage}%
    \hfill
    \begin{minipage}{0.49\textwidth}
        \includegraphics[width=\linewidth]{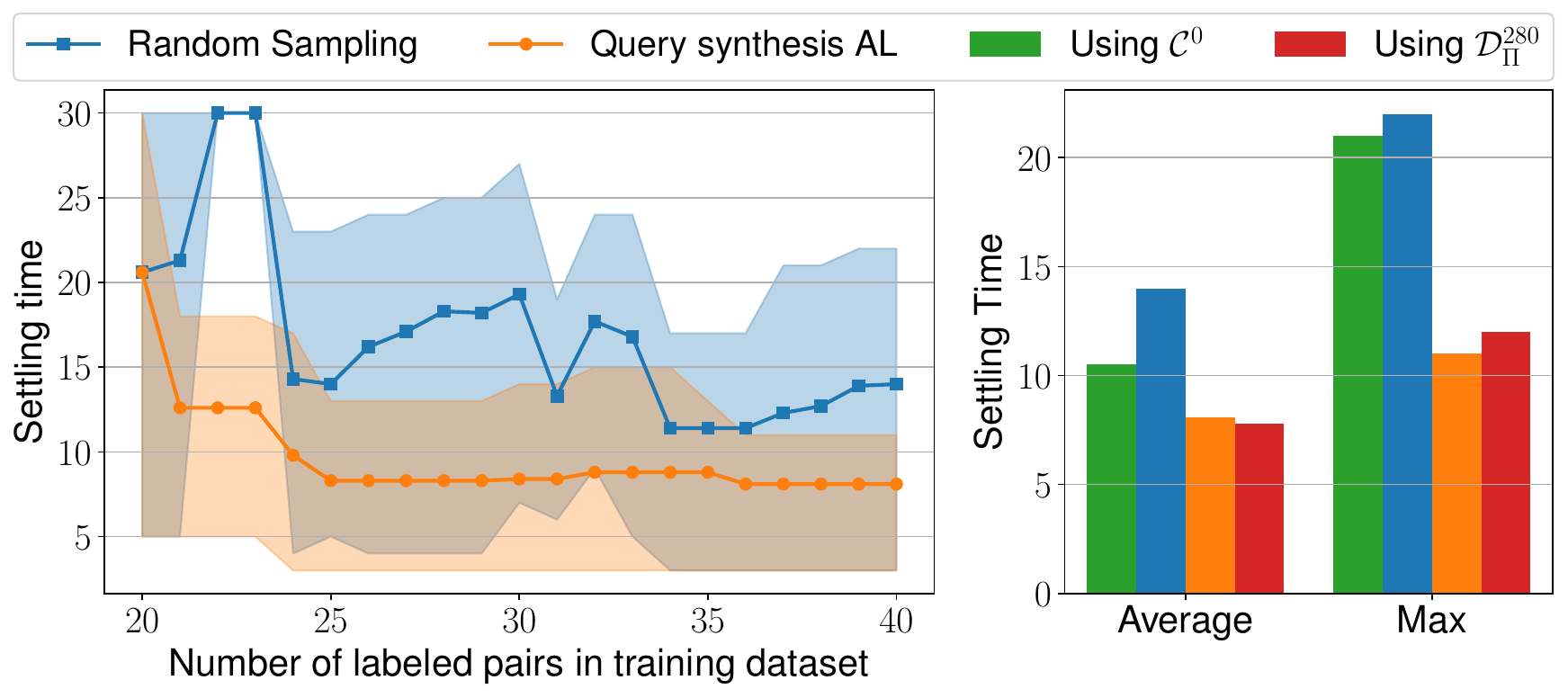}
    \end{minipage}
    \caption{Performance evaluation of active learning strategies: (a) pool-based AL using $D^k = D^k_p + d$, (b) query-synthesis AL. For each strategy, left plot shows settling time evolution during AL iterations with the shaded regions indicating the min/max and lines with markers indicating average; right plot shows final MPC performance after $20$ AL iterations, compared with random sampling and with the MPC learned using $\Pds^{280}$, i.e., after labeling all $\cc{Q}$.}
    \label{fig:AL_loop}
\end{figure*}

Fig.~\ref{fig:AL_loop} shows the results obtained from $20$ iterations of the proposed AL strategies, compared to random sampling, which follows Algorithm~\ref{alg:poolAL} but selects pairs randomly from $\mathcal{Q}$ instead of maximizing the acquisition function~\eqref{eq:acquisition}.
The results show that, using a dataset with $\np = 40$ labeled trajectory pairs, the proposed AL strategies outperform the MPC controllers in $\cc{C}^0$ and the MPC learned using random sampling.
Additionally, the AL strategies achieve a settling-time performance that is at least as good as the one obtained from the MPC learned from a large dataset $\Pds^{280}$, i.e., with $\np = 300$ ($20$ in $\Pds^0$ plus the $280$ from $\cc{Q}$).

\begin{figure}[t]
\centering
    \includegraphics[width=0.92\linewidth]{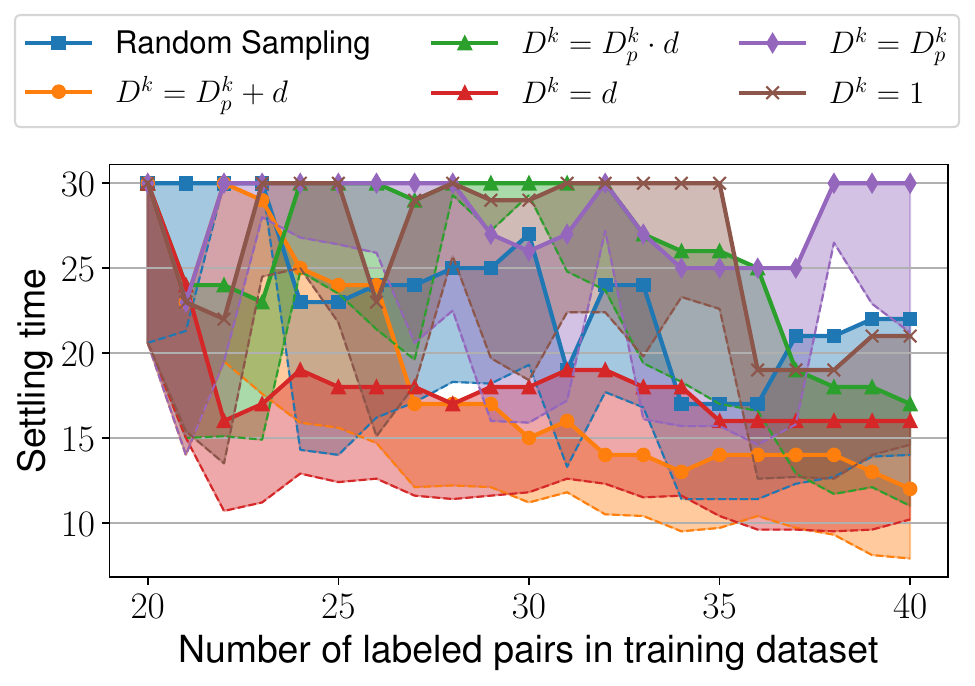}
    \caption{Settling times of pool-based AL using different $D^k$ in the acquisition function~\eqref{eq:acquisition}. Shaded areas indicate the region between maximum (solid lines with markers) and average values (dashes lines).}
    \label{fig:pb_dist}
\end{figure}

Fig.~\ref{fig:pb_dist} compares the performance of the pool-based AL strategy using different expressions of the diversity metric $D^k(\mathcal{P})$ in the acquisition function~\eqref{eq:acquisition}.
As mentioned in Section~\ref{sec:ALstrategies}, we find that combining $D^k_p$ and $d$ typically gives better results than only considering one of them, with $D^k(\mathcal{P}) = D^k_p(\mathcal{P}) + d(\mathcal{P})$ providing the best results in this case.
Note that the results using $D^k = 1$ only take into account the uncertainty $U(\cc{P})$.

Tables~\ref{tab:final_comp:pb} and~\ref{tab:final_comp:qs} shows further computational results for $200$ initial states $\hat{x} \in \cXo$, comparing each of the controllers obtained after $20$ AL iterations with the controller obtained using the full dataset $\Pds^{280}$, i.e., the original dataset $\Pds^{0}$ plus the $280$ samples from $\cc{Q}$.
Results for $\max \| \vv{u}(\hat{\traj}) \|_\infty$ only take into account the trajectories where the AL-learned MPC and $\Pds^{280}$ have the same settling time, as this is when the preference function~\eqref{eq:results:complex:prefFunc} takes this metric into account.
Pool-based MPC was preferred by~\eqref{eq:results:complex:prefFunc} over $\Pds^{280}$ $101$ out of the $200$ tested initial states ($35$ of these shared the same settling time), whereas query-synthesis MPC was preferred $98$ times ($61$ of these shared the same settling time).
The results show that the AL methods learn an MPC that is deemed by the human as good as the MPC obtained from a dataset with $300$ labeled pairings.

\section{Conclusions} \label{sec:conclusions}

We introduced two active learning strategies to reduce the number of queries required in the preference-based learning framework presented in \cite{Krupa_CDC_25}: a pool-based method that selects trajectory pairs from a fixed dataset using an acquisition function that combines uncertainty and trajectory diversity; and a query-synthesis method that generates new trajectory pairs during the learning process using the current learned MPC.
Numerical results have shown that the proposed AL strategies can outperform random sampling, achieving a performance that is closer to the one obtained by using a larger dataset of human queries.
Future research will explore the inclusion of an exploration term in the query-synthesis strategy and further investigate diversity metrics $D^k$.

\begin{ack}
This work was funded by the European Union (ERC Advanced Research Grant COMPACT, No. 101141351). Views
and opinions expressed are however those of the authors only and do not necessarily reflect those of the European Union
or the European Research Council. Neither the European Union nor the granting authority can be held responsible for
them.
\end{ack}

{\renewcommand{\arraystretch}{1.15}%
\begin{table}[t]
\setlength{\tabcolsep}{6.5pt}
\centering
\begin{tabular}{|c|cc|ccc|}
\hline
\multirow{2}*{MPC} & \multicolumn{2}{c|}{$\settime(\hat{\traj})$ } & \multicolumn{3}{c|}{Normalized $\max \| \vv{u}(\hat{\traj}) \|_\infty$} \\
\cline{2-6}
& Avrg. & Max. & Avrg. & Max. & Min. \\
\hline
Using $\Pds^{280} $ & 8.86 & 17 & 1.00 & 2.00 & 0.24 \\
Pool-based $\Pds^{20}$    & 8.48 & 13 & 1.10 & 2.00 & 0.22 \\
\hline
\end{tabular}
\caption{Pool-based comparison for 200 $\hat{x}$.}
\label{tab:final_comp:pb}
\end{table}
}

{\renewcommand{\arraystretch}{1.15}%
\begin{table}[t]
\setlength{\tabcolsep}{5pt}
\centering
\begin{tabular}{|c|cc|ccc|}
\hline
\multirow{2}*{MPC} & \multicolumn{2}{c|}{$\settime(\hat{\traj})$ } & \multicolumn{3}{c|}{Normalized $\max \| \vv{u}(\hat{\traj}) \|_\infty$} \\
\cline{2-6}
& Avrg. & Max. & Avrg. & Max. & Min. \\
\hline
Using $\Pds^{280} $ & 8.86 & 17 & 1.16 & 2.00 & 0.24 \\
Query-synth. $k {=} 20$ & 9.55 & 19 & 0.69 & 1.33 & 0.12 \\
\hline
\end{tabular}
\caption{Query-synth. comparison for 200 $\hat{x}$.}
\label{tab:final_comp:qs}
\end{table}
}

\bibliography{ifacconf}
\end{document}